\documentclass{pasj01}
\Received{2016 January 18}
\Accepted{2016 February 24}
\Published{}
\let\lsim=\lesssim
\let\gsim=\gtrsim
\begin{document}

\title{SXDF-ALMA 2 arcmin$^2$ Deep Survey: 1.1-mm number counts}

\author{
Bunyo~\textsc{Hatsukade},\altaffilmark{1*,$\dagger$}
Kotaro~\textsc{Kohno},\altaffilmark{2,3}
Hideki~\textsc{Umehata},\altaffilmark{2,4}
Itziar~\textsc{Aretxaga},\altaffilmark{5}
Karina~I.~\textsc{Caputi},\altaffilmark{6}
James~S.~\textsc{Dunlop},\altaffilmark{7}
Soh~\textsc{Ikarashi},\altaffilmark{6}
Daisuke~\textsc{Iono},\altaffilmark{1,8}
Rob~J.~\textsc{Ivison},\altaffilmark{4,7}
Minju~\textsc{Lee},\altaffilmark{1,9}
Ryu~\textsc{Makiya},\altaffilmark{2,10}
Yuichi~\textsc{Matsuda},\altaffilmark{1,8}
Kentaro~\textsc{Motohara},\altaffilmark{2}
Kouichiro~\textsc{Nakanishi},\altaffilmark{1,8}
Kouji~\textsc{Ohta},\altaffilmark{11}
Ken-ich~\textsc{Tadaki},\altaffilmark{12}
Yoichi~\textsc{Tamura},\altaffilmark{2}
Wei-Hao~\textsc{Wang},\altaffilmark{13}
Grant~W.~\textsc{Wilson},\altaffilmark{14}
Yuki~\textsc{Yamaguchi},\altaffilmark{2}
Min~S.~\textsc{Yun},\altaffilmark{14}
}
\altaffiltext{1}{National Astronomical Observatory of Japan, 2-21-1 Osawa, Mitaka, Tokyo 181-8588, Japan}
\email{bunyo.hatsukade@nao.ac.jp $^\dagger$NAOJ Fellow.}
\altaffiltext{2}{Institute of Astronomy, University of Tokyo, 2-21-1 Osawa, Mitaka, Tokyo 181-0015, Japan}
\altaffiltext{3}{Research Center for the Early Universe, The University of Tokyo, 7-3-1 Hongo, Bunkyo, Tokyo 113-0033, Japan}
\altaffiltext{4}{European Southern Observatory, Karl-Schwarzschild-Str. 2, D-85748 Garching, Germany}
\altaffiltext{5}{Instituto Nacional de Astrof\'{\i}sica, \'{O}ptica y Electr\'{o}nica (INAOE), Luis Enrique Erro 1, Sta. Ma. Tonantzintla, Puebla, Mexico}
\altaffiltext{6}{Kapteyn Astronomical Institute, University of Groningen, P.O. Box 800, 9700AV Groningen, The Netherlands}
\altaffiltext{7}{Institute for Astronomy, University of Edinburgh, Royal Observatory, Edinburgh EH9 3HJ UK}
\altaffiltext{8}{SOKENDAI (The Graduate University for Advanced Studies), 2-21-1 Osawa, Mitaka, Tokyo 181-8588, Japan}
\altaffiltext{9}{Department of Astronomy, The University of Tokyo, 7-3-1 Hongo, Bunkyo-ku, Tokyo 133-0033, Japan}
\altaffiltext{10}{Kavli Institute for the Physics and Mathematics of the Universe, Todai Institutes for Advanced Study, the University of Tokyo, Kashiwa, Japan 277-8583 (Kavli IPMU, WPI)}
\altaffiltext{11}{Department of Astronomy, Kyoto University, Kyoto 606-8502, Japan}
\altaffiltext{12}{Max-Planck-Institut f\"{u}r extraterrestrische Physik (MPE),Giessenbachstr., D-85748 Garching, Germany}
\altaffiltext{13}{Institute of Astronomy and Astrophysics, Academia Sinica, Taipei, Taiwan}
\altaffiltext{14}{Department of astronomy, University of Massachusetts, Amherst, MA 01003, USA}


\maketitle
%%%%%%%%%%%%%%%%%%%%%%%%%%%%%%%%%%%%%%%%%%%%%%%%%%%%%%%%%%%%%%%%%%%%%%%%%%%%%%%%
%%%%%%%%%%%%%%%%%%%%%%%%%%%%%%%%%%%%%%%%%%%%%%%%%%%%%%%%%%%%%%%%%%%%%%%%%%%%%%%%
\begin{abstract}
We report 1.1~mm number counts revealed with the Atacama Large Millimeter/submillimeter Array (ALMA) in the Subaru/XMM-Newton Deep Survey Field (SXDF). 
The advent of ALMA enables us to reveal millimeter-wavelength number counts down to the faint end without source confusion. 
However, previous studies are based on the ensemble of serendipitously-detected sources in fields originally targeting different sources and could be biased due to the clustering of sources around the targets. 
We derive number counts in the flux range of 0.2--2~mJy by using 23 ($\ge$4$\sigma$) sources detected in a continuous 2.0-arcmin$^2$ area of the SXDF. 
The number counts are consistent with previous results within errors, suggesting that the counts derived from serendipitously-detected sources are not significantly biased, although there could be field-to-field variation due to the small survey area. 
By using the best-fit function of the number counts, we find that $\sim$40\% of the extragalactic background light at 1.1~mm is resolved at $S_{\rm 1.1mm} > 0.2$~mJy. 
\end{abstract}

%%%%%%%%%%%%%%%%%%%%%%%%%%%%%%%%%%%%%%%%%%%%%%%%%%%%%%%%%%%%%%%%%%%%%%%%%%%%%%%%
%%%%%%%%%%%%%%%%%%%%%%%%%%%%%%%%%%%%%%%%%%%%%%%%%%%%%%%%%%%%%%%%%%%%%%%%%%%%%%%%
\section{Introduction}

Deep and wide-field surveys discovered a population of galaxies bright at millimeter/submillimeter (mm/submm) wavelengths (SMGs). 
SMGs are dusty starburst galaxies at high redshifts with star-formation rates (SFRs) of a few 100--1000~$M_{\odot}$~yr$^{-1}$, and hold important clues to the true star formation history and the galaxy evolution in the universe (e.g., \cite{blai02, case14}). 
SMGs are also important for understanding the origin of extragalactic background light (EBL), which is thought to be the integral of unresolved emission from extragalactic sources. 
While the EBL at mm/submm is thought to be largely contributed by distant dusty galaxies (e.g., \cite{laga05}), the contribution of bright SMGs ($S_{\rm 1mm} > 1$~mJy) detected in previous 1-mm blank field surveys is $\lsim$20\% (e.g., \cite{grev04, scot08, scot10, hats11}), suggesting that the rest of the EBL originates from `sub-mJy' sources. 
Atacama Large Millimeter/submillimeter Array (ALMA) enables us to explore the flux regime more than an order of magnitude fainter than those detected in previous single-dish surveys because of its high sensitivity and high angular resolution. 
ALMA has probed the faint end of the number counts and more than half of the EBL has been resolved in previous studies (\cite{hats13, ono14, carn15, fuji16}). 
However, these studies are based on the ensemble of serendipitously-detected sources in fields originally targeting different sources, and the number counts obtained in those fields could be biased due to the clustering of sources around the targets or sidelobes caused by bright targets.

In this paper, we present 1.1~mm number counts in a contiguous area revealed by the ALMA observations of a part of the Subaru/{\sl XMM-Newton} Deep Survey Field (SXDF) \citep{furu08} which includes the field of the Cosmic Assembly Near-IR Deep Extragalactic Legacy Survey (CANDELS; \cite{grog11, koek11}). 
The survey design and source catalog is described in Kohno et al. (in prep.), and the properties of detected sources and other galaxy populations in the field are discussed in \cite{tada15} and subsequent papers. 
The arrangement of this paper is as follows. 
Section~\ref{sec:data} outlines the data we used, and Section~\ref{sec:source} describes the method of source extraction and simulations carried out to estimate the number counts. 
In Section~\ref{sec:discussion}, we compare our derived number counts with other observational results and model predictions, and estimate the contribution of the 1.1~mm sources to the EBL.

%%%%%%%%%%%%%%%%%%%%%%%%%%%%%%%%%%%%%%%%%%%%%%%%%%%%%%%%%%%%%%%%%%%%%%%%%%%%%%%%
%%%%%%%%%%%%%%%%%%%%%%%%%%%%%%%%%%%%%%%%%%%%%%%%%%%%%%%%%%%%%%%%%%%%%%%%%%%%%%%%
\begin{figure}
\begin{center}
\includegraphics[width=\linewidth]{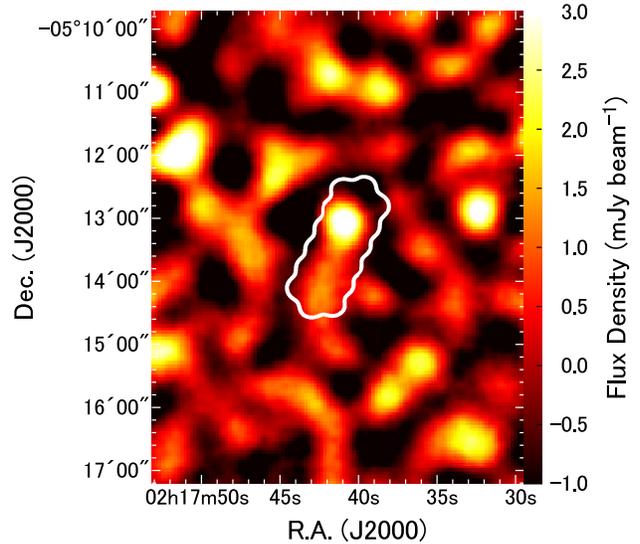}
\end{center}
\caption{
AzTEC 1.1~mm map around the SXDF-ALMA survey field. 
The white curve shows the 50\% coverage region used in this study. 
}
\label{fig:aztec}
\end{figure}

%===============================================================================
\begin{figure*}
\begin{center}
\includegraphics[width=\linewidth]{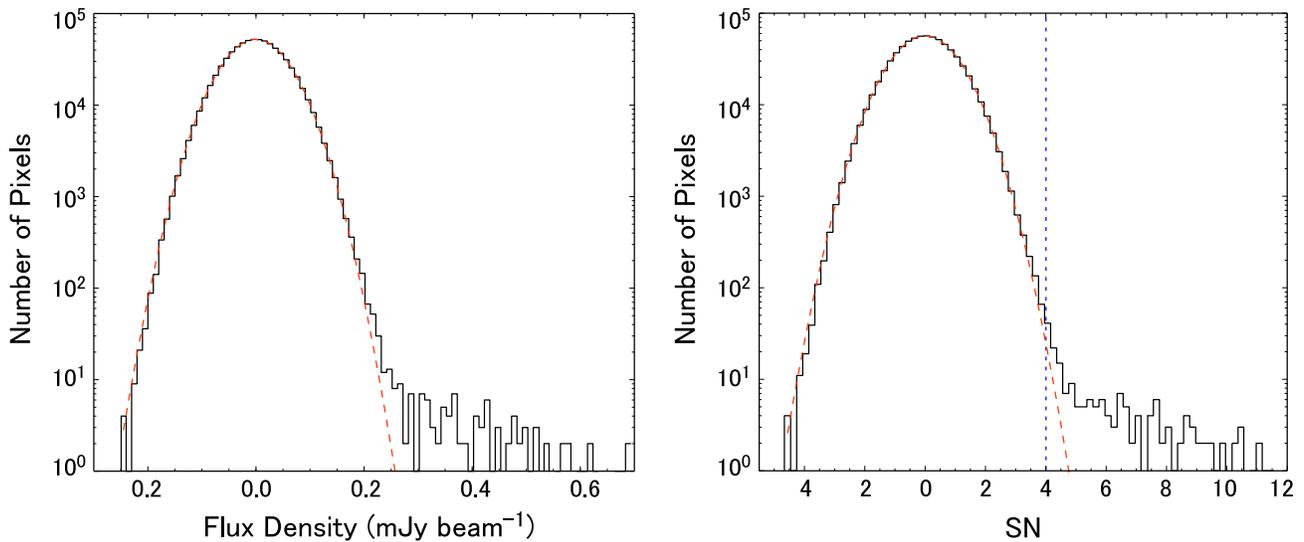}
\end{center}
\caption{
Distribution of flux density of the signal map (left) and SN (right) within the 50\% coverage region. 
The dashed curves show the result of a Gaussian fit. 
The vertical dotted line in the right panel indicates the SN threshold ($4\sigma$) for source extraction used in this study. 
\label{fig:pix_hist}}
\end{figure*}

%===============================================================================
\section{SXDF-ALMA Survey Data}\label{sec:data}

The details of the ALMA observations (Program ID: 2012.1.00756.S; PI: K. Kohno) and data reduction are described in Kohno et al. (in prep.) and \citet{kohn16}, and here we briefly summarize them. 
We conducted band 6 (1.1~mm, or 274~GHz) imaging of a contiguous area in the SXDF during the ALMA Cycle 1 session. 
The observing field is selected to cover a 1.1-mm source detected with AzTEC ($S_{\rm 1.1mm} = 3.5^{+0.6}_{-0.5}$~mJy; Ikarashi et al. in prep.) and 12 H$\alpha$-selected star-forming galaxies (figure~\ref{fig:aztec}, see also \cite{tada15}). 
The field was set to $105'' \times 50''$ in the the ALMA Observing Tool and has an effective area of 2.0~arcmin$^2$ (``50\% coverage region'' defined below). 
The field was covered by a 19-point mosaic with a total observing time of 3.6 hours. 
The number of available antennas was 30--32 and the range of baseline lengths was 20--650~m. 
The correlator was used in the time domain mode with a bandwidth of 1875 MHz, and the total bandwidth with four spectral windows is 7.5~GHz. 
The data were reduced with Common Astronomy Software Applications (CASA; \cite{mcmu07}). 
The map was processed with {\verb CLEAN } algorithm with the natural weighting, which gives the synthesized beamsize of $0\farcs53 \times 0\farcs41$ (position angle of 63\fdg9). 
In this study, we use the region where the primary beam attenuation is less than or equal to 50\% in the map (``50\% coverage region''), which is a 2.0-arcmin$^2$ area. 
A sensitivity map was created by using the {\sc aips} \citep{grei03} {\sc rmsd} task with a box size of 100 pixel $\times$ 100 pixel ($10'' \times 10''$) from the map before primary beam correction. 
The range of rms noise level within the 50\% coverage region is 48--61~$\mu$Jy~beam$^{-1}$ and the typical rms noise level is 55~$\mu$Jy~beam$^{-1}$. 
Figure~\ref{fig:pix_hist} shows the distributions of flux density of the signal map and signal-to-noise ratio (SN) within the 50\% coverage region. 
The SN distribution is created from the signal map divided by the sensitivity map. 
The vertical dotted line in the panel of SN distribution indicates the threshold for source extraction adopted in this study ($4\sigma$). 
The distributions are well explained by a Gaussian and a Gaussian fit to the flux distribution gives $1 \sigma$ of 55~$\mu$Jy~beam$^{-1}$. 
The excess from the fitted Gaussian at $S_{\rm 1.1mm} \gsim 0.2$~mJy in the flux distribution or at SN $\gsim$ $4\sigma$ in the SN distribution indicates the contribution from real sources.

%%%%%%%%%%%%%%%%%%%%%%%%%%%%%%%%%%%%%%%%%%%%%%%%%%%%%%%%%%%%%%%%%%%%%%%%%%%%%%%%
%%%%%%%%%%%%%%%%%%%%%%%%%%%%%%%%%%%%%%%%%%%%%%%%%%%%%%%%%%%%%%%%%%%%%%%%%%%%%%%%
\section{Source Detection and Number Counts}\label{sec:source}

%%%%%%%%%%%%%%%%%%%%%%%%%%%%%%%%%%%%%%%%%%%%%%%%%%%%%%%%%%%%%%%%%%%%%%%%%%%%%%%%
\subsection{Source and Spurious Detection}\label{sec:spurious}

Source detection was conducted on the image before correcting for the primary beam attenuation. 
We adopt a source-finding algorithm {\sc aegean} \citep{hanc12}, which achieves high reliability and completeness performance for radio maps compared to other source-finding packages 
and is used in radio or submm surveys (e.g., \cite{umeh15}). 
We find 25 (6) sources with a peak SN of $\ge$$4\sigma$ ($\ge$$5\sigma$). 
The range of peak flux density of the $\ge$$4\sigma$ sources is 0.2--1.7 mJy after primary beam correction. 
In this study we use the peak flux densities for creating number counts. 
We check that the peak flux densities ($S_{\rm peak}$) are consistent with integrated flux densities ($S_{\rm integ}$) measured with the {\sc aips}/{\sc sad} task within errors with the ratio of $S_{\rm integ}/S_{\rm peak} = 1.1 \pm 0.3$ (see Kohno et al. in prep.).

It is possible that the SXDF-ALMA field is overdense because the field is selected to include an AzTEC source (figure~\ref{fig:aztec}). 
The number density of AzTEC sources in the SXDF-ALMA field is 0.5 arcmin$^{-2}$, which is a factor of 1.4 higher than that of the original AzTEC 1.1~mm survey of 0.36 arcmin$^{-2}$ \citep{scot12}. 
In the ALMA survey, we found that the two brightest ALMA sources are associated with the AzTEC source (Kohno et al. in prep.; Yamaguchi et al. in prep.). 
In what follows, we exclude the two ALMA sources when deriving number counts to avoid the effect of the possible overdensity.

\begin{figure}
\begin{center}
\includegraphics[width=\linewidth]{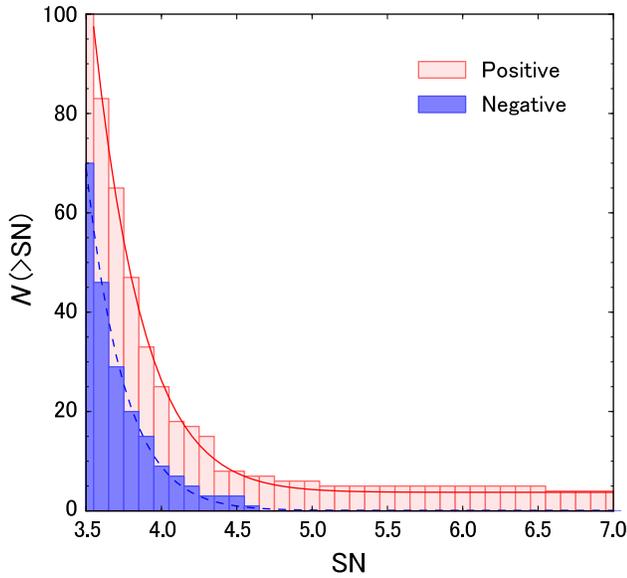}
\end{center}
\caption{
Cumulative number of positive and negative peaks as a function of SN threshold. 
Solid and dashed curves represent the best-fit function of $f({\rm SN}) = A [1 - {\rm erf}({\rm SN}/B)]+C$ for the positive peaks and negative peaks. 
}
\label{fig:fdr}
\end{figure}

In order to estimate the degree of contamination by spurious sources, we count the number of negative peaks as a function of SN threshold (figure~\ref{fig:fdr}). 
Nine negative sources are found at $\ge$$4\sigma$, and no negative source at $\ge$$4.7\sigma$. 
The probability of contamination by spurious sources is estimated from the fraction of negative peaks to positive peaks as a function of SN and is considered when creating number counts.

%===============================================================================

%%%%%%%%%%%%%%%%%%%%%%%%%%%%%%%%%%%%%%%%%%%%%%%%%%%%%%%%%%%%%%%%%%%%%%%%%%%%%%%%
\subsection{Completeness}\label{sec:completeness}
\begin{figure}
\begin{center}
\includegraphics[width=\linewidth]{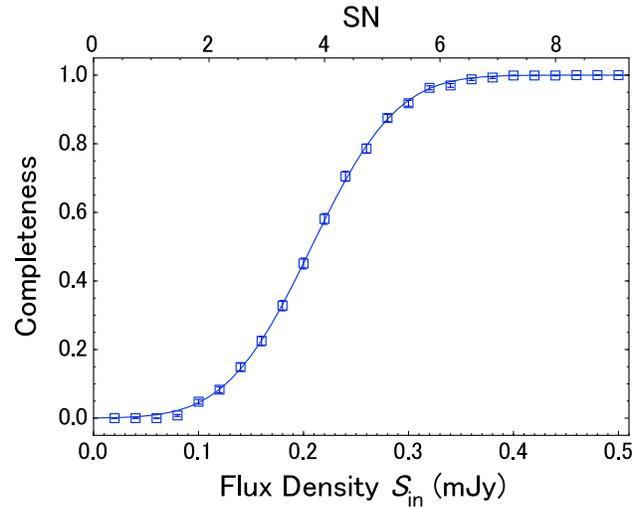}
\end{center}
\caption{
Completeness as a function of input flux ($S_{\rm in}$) (corrected for primary beam attenuation). 
The error bars are 1$\sigma$ from the binomial distribution. 
The top axis shows the effective SN by using a typical rms noise level of 55~$\mu$Jy~beam$^{-1}$. 
Solid curve represents the best-fit function of $f(S_{\rm in}) = [1 +  {\rm erf}((S_{\rm in} - A)/B)]/2$. 
}
\label{fig:completeness}
\end{figure}

We calculate the completeness, which is the rate at which a source is expected to be detected in a map, to see the effect of noise fluctuations on the source detection. 
The completeness calculation is conducted on the map corrected for primary beam attenuation. 
An artificial source made by scaling the synthesized beam is injected into a position randomly selected in the map $>$$1.0''$ away from $\ge$$3\sigma$ peaks to avoid the contribution from nearby sources. 
We checked that the completeness does not change significantly if we remove the constraint on the input source positions. 
Note that the nearest peak of sidelobes of the synthesized beam is located $1''$ away from the center and the relative flux is less than 6\% of the main beam, and the effect of the sidelobe is negligible. 
When the input source is extracted within $1.0''$ of its input position with $\ge$$4\sigma$, the source is considered to be recovered. 
This procedure is repeated 1000 times for each flux bin. 
The result is shown in figure~\ref{fig:completeness}, and the completeness in the flux range of the detected sources is $\sim$50\%--100\%. 
We confirmed that the completeness and resulting number counts do not depend sensitively on the SN threshold cut for the completeness simulations.

\begin{figure}
\begin{center}
\includegraphics[width=\linewidth]{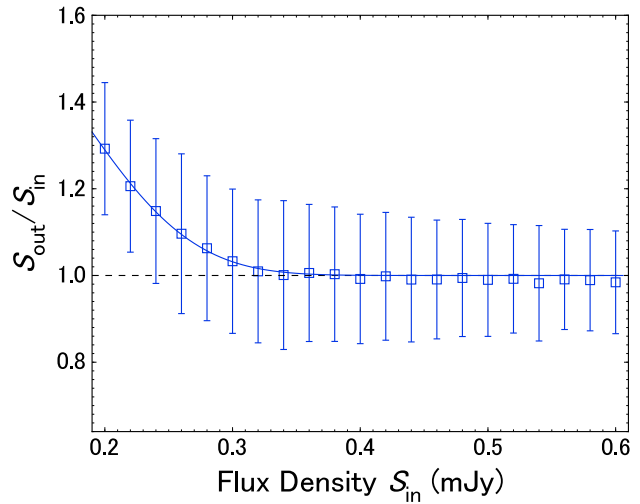}
\end{center}
\caption{
Ratio between input fluxes ($S_{\rm in}$) and output fluxes ($S_{\rm out}$) as a function of input flux (corrected for primary beam attenuation). 
Error bars show $1\sigma$ of 1000 trials. 
Solid curve represents the best-fit function of $f(S_{\rm in}) = 1 + A \exp(-B S_{\rm in}^C)$. 
}
\label{fig:boosting}
\end{figure}

When dealing with a low SN map, we need to consider the effect that flux densities of low SN sources are boosted (\cite{murd73, hogg98}). 
In the course of this simulation, we calculate the ratio between the input and output fluxes to estimate the intrinsic flux density of the detected sources (figure~\ref{fig:boosting}). 
The ratio for the flux range of the detected sources is 1.0--1.3.

%===============================================================================

%===============================================================================

%%%%%%%%%%%%%%%%%%%%%%%%%%%%%%%%%%%%%%%%%%%%%%%%%%%%%%%%%%%%%%%%%%%%%%%%%%%%%%%%
%%%%%%%%%%%%%%%%%%%%%%%%%%%%%%%%%%%%%%%%%%%%%%%%%%%%%%%%%%%%%%%%%%%%%%%%%%%%%%%%
\subsection{Number Counts}\label{sec:counts}

\begin{figure*}
\begin{center}
\includegraphics[width=\linewidth]{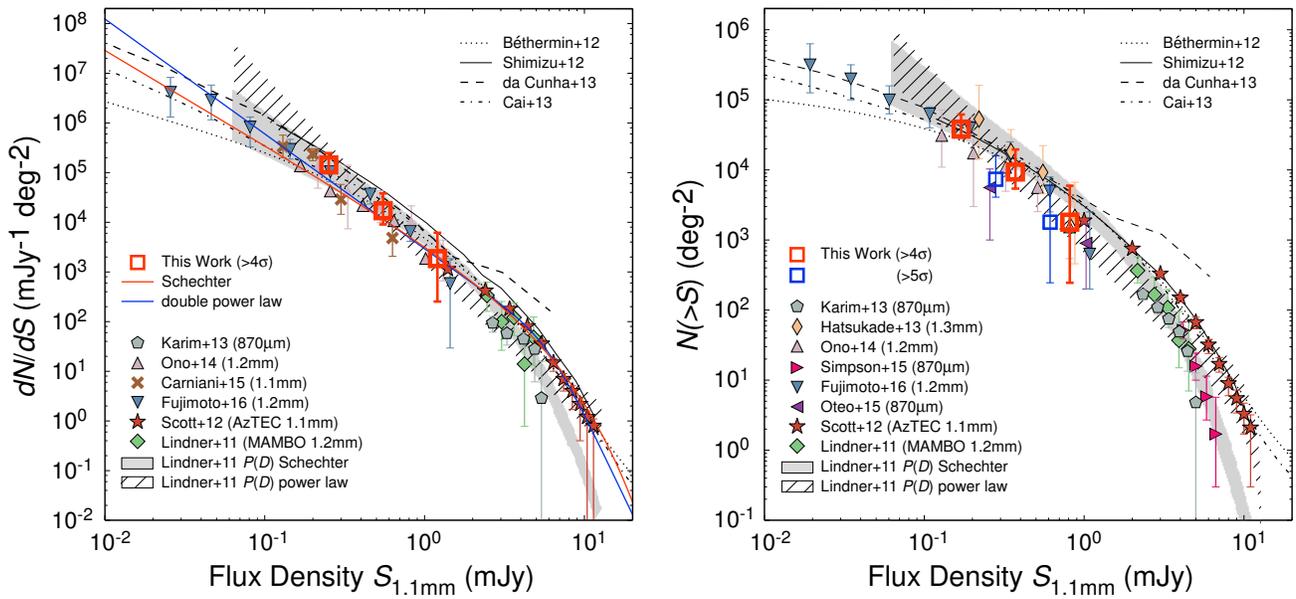}
\end{center}
\caption{
Differential (left) and cumulative (right) number counts at 1.1~mm obtained in the SXDF-ALMA survey (red squares). 
In addition to the counts for $\ge$$4\sigma$ sources, the cumulative counts for $\ge$$5\sigma$ sources (blue squares) are also presented. 
For comparison, we plot ALMA number counts in the ECDFS \citep{kari13} and the UKIDSS UDS field \citep{simp15b}, 
and counts derived from ALMA-detected serendipitous sources by \citet{hats13}, \citet{ono14}, \citet{carn15}, \citet{fuji16}, and \citet{oteo15}. 
Number counts derived with single-dish surveys with MAMBO \citep{lind11} (Lockman Hole North field) and AzTEC \citep{scot12} (combined counts of 6 blank fields) are also presented.
Shaded and hatched regions show 95\% confidence regions for the best-fit Schechter function and power-law function models derived from the {\it P(D)} analysis by \citet{lind11}. 
Red and blue solid curves represent the best-fit functions of Schechter function and double power law function, respectively (see Section~\ref{sec:discussion}).
We also show model predictions of \citet{shim12}, \citet{beth12}, \citet{cai13}, and \citet{dacu13}. %citet{hayw13}, 
The flux density of the counts are scaled to 1.1~mm flux density by assuming a modified black body with typical values for SMGs (spectral index of $\beta = 1.5$, dust temperature of 35~K (e.g., \cite{kova06, copp08}), and $z = 2.5$): $S_{\rm 1.1 mm}/S_{\rm 870 \mu m} = 0.56$, $S_{\rm 1.1 mm}/S_{\rm 1.2 mm} = 1.29$, and $S_{\rm 1.1 mm}/S_{\rm 1.3 mm} = 1.48$. 
}
\label{fig:counts}
\end{figure*}

By using the $\ge$$4 \sigma$ sources, we create differential and cumulative number counts. 
To create number counts, we correct for the contamination of spurious sources, the completeness, and the flux boosting. 
The contamination of spurious sources to each source is estimated as a fraction of the number of positive peaks to negative peaks at its SN by using the best-fit functions (figure~\ref{fig:fdr}) and is subtracted from unity. 
Then the counts are divided by the completeness by using the best-fit function (figure~\ref{fig:completeness}). 
The flux boosting is corrected by using the best-fit function shown in figure~\ref{fig:boosting}. 
The errors are calculated by accounting for Poisson fluctuations, uncertainties on corrections for completeness and for flux boosting, and field-to-field variation. 
Contributions from the uncertainties are calculated separately and combined by taking the square root of the sum of the squares in each flux bin. 
The uncertainties from Poisson fluctuations is estimated from Poisson confidence limits of 84.13\% \citep{gehr86}, which correspond to $1 \sigma$ for Gaussian statistics that can be applied to small number statistics. 
The errors due to the uncertainties on corrections for completeness and flux boosting are estimated by using their $1 \sigma$ errors (figure~\ref{fig:completeness} and \ref{fig:boosting}). 
These errors are calculated for each source and propagated to the errors of each flux bin. 
The uncertainties from field-to-field variation are estimated by using a publicly available tool of \citet{tren08}. 
We use the survey area of $2' \times 1'$, the mean redshift of $\bar{z}=2.5$, the redshift range of $\Delta z = 3$, which are derived for SMGs (e.g., \cite{chap05, yun12}), and a halo occupation fraction of unity. 
%It uses , mean redshift ($\bar{z}$), redshift range ($\Delta z$), and the number of sources, and we assume $\bar{z}=2.5$ and $\Delta z = 3$, which are derived for SMGs (e.g., \cite{chap05, yun12}). 
We adopted default cosmological parameters of $\Omega_{\rm{M}} = 0.26$, $\Omega_{\Lambda} = 0.74$, $h = 70$ km s$^{-1}$ Mpc$^{-1}$ \citep{sper07}, and $\sigma_8 = 0.9$. 
The bias calculation is conducted by using the formalism of \cite{shet99}. 
As a result, we find that the relative errors from field-to-field variation is 10\%--20\% for the bins of the differential counts. 
The derived number counts and errors are shown in figure~\ref{fig:counts} and Table~\ref{tab:counts}. 
The total errors are dominated by the contribution from Poisson fluctuations.

The differential number counts obtained in this study and previous studies are fitted to a Schechter function of the form, $dN/dS = N'/S' (S/S')^{\alpha} \exp(-S/S')$, 
and a double power-law function of the form, $dN/dS = N'/S' [ (S/S')^{\alpha} + (S/S')^{\beta} ]^{-1}$. 
In this fit, we use the number counts obtained in 1-mm observations (1.1--1.3~mm) with ALMA and single-dish telescopes plotted in figure~\ref{fig:counts} (left) but do not use the number counts at 870~$\mu$m to avoid the uncertainties from flux scaling. 
We also do not use the faint end of number counts ($S < 0.1$~mJy) by \citet{fuji16}, where the counts are derived by including gravitationally lensed sources to avoid uncertainty from the magnification correction. 
The best-fit parameters are summarized in Table~\ref{tab:fit}. 
The power law slopes are consistent with those of \citet{fuji16}.

%===============================================================================

%===============================================================================
\begin{table}
\tbl{Differential and cumulative number counts. \label{tab:counts}}
{
\footnotesize
\begin{tabular}{cccccc}
\hline
\multicolumn{3}{c}{Differential Counts} & \multicolumn{3}{c}{Cumulative Counts} \\
$S$ & $N$ & $dN/dS$ & $S$ & $N$ & $N(>S)$ \\
(mJy) & & ($10^3$ mJy$^{-1}$ deg$^{-2}$) & (mJy) & & ($10^3$ deg$^{-2}$) \\
\hline
0.25 & 17 & $143.2^{+104.2}_{-41.0}$ & 0.17 & 23 & $38.2^{+23.5}_{-9.4}$ \\
0.55 &  5 & $ 17.1^{+21.1}_{-8.0}$   & 0.37 &  6 & $ 9.3^{+10.2}_{-3.9}$ \\
1.20 &  1 & $  1.9^{+4.3}_{-1.6}$    & 0.81 &  1 & $ 1.8^{+4.2}_{-1.6}$ \\
\hline
\end{tabular}}
\begin{tabnote}
The errors are $1\sigma$. 
\end{tabnote}
\end{table}

%===============================================================================
\begin{table}
\tbl{Best-fit parameters of parametric fits to differential number counts. \label{tab:fit}}
{
\footnotesize
\begin{tabular}{ccccc}
\hline
Function & $N'$ & $S'$ & $\alpha$ & $\beta$ \\
         & ($10^2$ deg$^{-2}$) & (mJy) & &  \\
\hline
Schechter        & $15.0\pm 5.1$ & $3.1 \pm 0.5$ & $-1.9 \pm 0.2$ & $-$ \\
Double power law & $2.9 \pm 0.6$ & $5.9 \pm 0.5$ & $ 6.8 \pm 0.8$ & $2.3 \pm 0.1$ \\
\hline
\end{tabular}}
\begin{tabnote}
The errors are $1\sigma$. 
\end{tabnote}
\end{table}

%%%%%%%%%%%%%%%%%%%%%%%%%%%%%%%%%%%%%%%%%%%%%%%%%%%%%%%%%%%%%%%%%%%%%%%%%%%%%%%%
%%%%%%%%%%%%%%%%%%%%%%%%%%%%%%%%%%%%%%%%%%%%%%%%%%%%%%%%%%%%%%%%%%%%%%%%%%%%%%%%
\section{Discussion and Conclusions}\label{sec:discussion}
We compare our SXDF-ALMA number counts with previous results. 
Number counts at faint flux densities ($\lsim$1~mJy) have been obtained by using serendipitously-detected sources within the field of view of original targets in each project \citep{hats13,ono14,carn15,fuji16}. 
These counts could be biased because of the possible clustering around the original targets. 
In figure~\ref{fig:counts}, we plot ALMA number counts obtained in the Extended Chandra Deep Field South (ECDFS) by \citet{kari13} (870~$\mu$m) and in the UKIRT InfraRed Deep Sky Surveys (UKIDSS) Ultra Deep Survey (UDS) field by \citet{simp15b} (870~$\mu$m), 
and ALMA number counts derived from serendipitously-detected sources by \citet{hats13} (1.3~mm), \citet{ono14} (1.2~mm), \citet{carn15} (1.1~mm), \citet{fuji16} (1.2~mm), and \citet{oteo15} (1.2~mm). 
We also show the number counts obtained by single-dish surveys with Max-Planck millimeter bolometer (MAMBO) at 1.2~mm \citep{lind11} and AzTEC at 1.1~mm \citep{scot12} for the bright end. 
The flux density of these counts are scaled to 1.1~mm flux density (see caption of figure~\ref{fig:counts}). 
The SXDF-ALMA counts obtained in a continuous region are consistent with previous results within errors, suggesting that the counts derived from serendipitous sources are not significantly biased, although our survey area is small and could be affected by field-to-field variation.

\citet{lind11} performed a fluctuation analysis (or {\it P(D)} analysis) of a single-dish map (beam FWHM of $11''$), which enables a model-dependent estimation of the number counts at faint flux densities below the confusion limit by using the information of the pixel flux distribution of a signal map rather than using detected sources. 
They assumed two types of function models for parameterization in the {\it P(D)} analysis; a single power law model and a Schechter function model. 
The number counts derived from the {\it P(D)} analysis are shown in figure~\ref{fig:counts}. 
The SXDF-ALMA counts are consistent with both functions within errors, but they are more consistent with the Schechter function {\it P(D)} counts than the single power law counts (figure~\ref{fig:counts}). 
This consistency suggests the validity of the {\it P(D)} analysis below the confusion limit or nominal sensitivity although the {\it P(D)} analysis is model dependent.

Recently \citet{oteo15} present the number counts by using $\ge$$5\sigma$ sources detected in the ALMA archival data of calibrators. 
Their counts are lower than previous results by a factor of at least two. 
They argue that the lower SN threshold ($<$$5\sigma$) adopted in the previous studies might include spurious sources. 
In order to verify this possibility, we create number counts by using only $\ge$$5\sigma$ sources, where no contamination of spurious sources are expected (Section~\ref{sec:spurious}). 
The resultant cumulative counts (figure~\ref{fig:counts}, right) decrease, but they are still consistent with the counts for $\ge$$4\sigma$ sources within the errors. 
A larger number of sources are needed to accurately constrain the number counts and verify the effect of spurious sources. 
\citet{oteo15} also argue that their counts are free from field-to-field variation because they collect data sets in different sky positions. 
Because we use sources detected in a limited survey area, it is possible that the field-to-field variation affects the number counts.

In figure~\ref{fig:counts}, recent model predictions with different approaches by \citet{shim12}, \citet{beth12}, \citet{cai13}, and \citet{dacu13} are also compared. 
\citet{shim12} perform large cosmological hydrodynamic simulations and simulate the properties of SMGs by calculating the reprocessing of stellar light by dust grains into far-IR to millimeter wavelengths in a self-consistent manner. 
The model of \citet{beth12} is based on the redshift evolution of the mass function of star-forming galaxies, specific SFR distribution at fixed stellar mass, and spectral energy distributions (SEDs) for the two star-formation modes (main-sequence and starburst). 
\citet{dacu13} create self-consistent models of the observed optical/NIR SEDs of galaxies detected in the Hubble Ultra Deep Field. 
They combine the attenuated stellar spectra with a library of IR emission models, which are consistent with the observed optical/NIR emission in terms of energy balance, and estimate the continuum flux at mm and submm wavelengths. 
The model of \citet{cai13} is a hybrid approach which combines a physical forward model for spheroidal galaxies and the early evolution of AGNs with a phenomenological backward model for late-type galaxies and the later evolution of AGNs. 
We found that the model predictions agree with our number counts within the uncertainties from both the models and the observed counts. 
In order to discern the models, it is important to probe fainter flux densities ($S_{\rm 1mm} \lsim 0.1$~mJy), where the difference among the models increases.

%%%%%%%%%%%%%%%%%%%%%%%%%%%%%%%%%%%%%%%%%%%%%%%%%%%%%%%%%%%%%%%%%%%%%%%%%%%%%%%%
%\subsection{Contribution to the Extragalactic Background Light}

\begin{figure}
\begin{center}
\includegraphics[width=\linewidth]{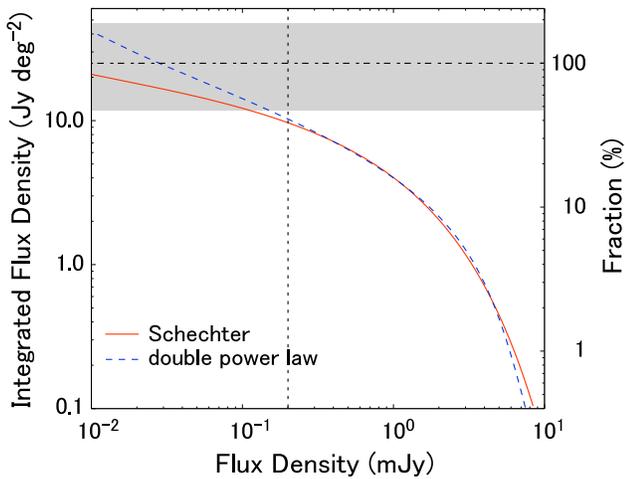}
\end{center}
\caption{
Integrated flux density of the best-fit functions (Schechter and double power law) of the differential number counts as a function of 1.1~mm flux. 
The right axis indicates the percentage of the resolved EBL. 
The horizontal dashed-dotted line and the shaded region show the EBL and its errors measured by \citet{fixs98}. 
The vertical dotted line indicates the flux limit in this study ($S_{\rm 1.1mm} > 0.2$). 
}
\label{fig:ebl}
\end{figure}

In this study, we derived 1.1~mm number counts in a continuous 2.0~arcmin$^2$ area of the SXDF-ALMA survey field. 
By using the best-fit functions of the number counts, we calculate the fraction of the 1.1~mm EBL revolved into discrete sources. 
The EBL at 1.1~mm based on the measurements by the {\sl Cosmic Background Explorer} satellite is $\sim$18 Jy~deg$^{-2}$ \citep{puge96} and $25^{+23}_{-13}$ Jy~deg$^{-2}$ \citep{fixs98}.
Figure~\ref{fig:ebl} shows the integrated flux density of the best-fit functions with a Schechter function and a double power law function derived from the differential number counts. 
The fraction of resolved EBL in this study is about 40\% at $S_{\rm 1.1mm} > 0.2$~mJy, which is consistent with previous studies with ALMA \citep{hats13, ono14, carn15, fuji16}. 
The integration of the best-fitting functions reaches 100\% at $S_{\rm 1.1mm} \sim 0.01$--0.03~mJy, although there is a large uncertainty to extend the functions to the fainter flux densities. 
In order to fully understand the origin of the EBL and to constrain the number counts without the effect of small number statistics and field-to-field variation, deeper ($S_{\rm 1.1mm} < 0.1$~mJy) and wider-area surveys in blank fields are essential in future studies.

%===============================================================================

%%%%%%%%%%%%%%%%%%%%%%%%%%%%%%%%%%%%%%%%%%%%%%%%%%%%%%%%%%%%%%%%%%%%%%%%%%%%%%%%
%%%%%%%%%%%%%%%%%%%%%%%%%%%%%%%%%%%%%%%%%%%%%%%%%%%%%%%%%%%%%%%%%%%%%%%%%%%%%%%%
\begin{ack}

We are grateful to Andrew J. Baker for pointing out the importance of fluctuation analysis and Robert Lindner for providing their number counts. 
We thank the referee for helpful comments and suggestions.
BH, YT, and and YM is supported by Japan Society for Promotion of Science (JSPS) KAKENHI (Nos.\ 15K17616, 25103503, 15H02073, 20647268).
KK is supported by the ALMA Japan Research Grant of NAOJ Chile Observatory, NAOJ-ALMA-0049. 
KC and SI acknowledge the support of the Netherlands Organisation for Scientific Research (NWO) through the Top Grant Project 614.001.403. 
JSD acknowledges the support of the European Research Council via an Advanced Grant. 
ML is financially supported by a Research Fellowship from JSPS for Young Scientists. 
HU acknowledges the support from Grant-in-Aid for JSPS Fellows, 26.11481.
This paper makes use of the following ALMA data: ADS/JAO.ALMA\#2012.1.00756.S. 
ALMA is a partnership of ESO (representing its member states), NSF (USA) and NINS (Japan), together with NRC (Canada), NSC and ASIAA (Taiwan), and KASI (Republic of Korea), in cooperation with the Republic of Chile. The Joint ALMA Observatory is operated by ESO, AUI/NRAO and NAOJ.

\end{ack}

%%%%%%%%%%%%%%%%%%%%%%%%%%%%%%%%%%%%%%%%%%%%%%%%%%%%%%%%%%%%%%%%%%%%%%%%%%%%%%%%
%%%%%%%%%%%%%%%%%%%%%%%%%%%%%%%%%%%%%%%%%%%%%%%%%%%%%%%%%%%%%%%%%%%%%%%%%%%%%%%%

%%%%%%%%%%%%%%%%%%%%%%%%%%%%%%%%%%%%%%%%%%%%%%%%%%%%%%%%%%%%%%%%%%%%%%%%%%%%%%%%
\end{document}